\documentclass[twocolumn]{IEEEtran}

\usepackage{booktabs}
\usepackage{url}
\usepackage{caption}
\usepackage{subcaption}
\usepackage{graphicx}
\usepackage{hyperref}
\hypersetup{
    colorlinks=true,
    linkcolor=blue,
    filecolor=magenta,      
    urlcolor=blue,
    citecolor=cyan
    }
\usepackage[utf8]{inputenc}
\usepackage[english]{babel}
\usepackage{csquotes}
\usepackage{graphicx}
\usepackage{color}
\usepackage{url}
\usepackage{subfiles}
\usepackage{authblk}

\usepackage[normalem]{ulem}
\useunder{\uline}{\ul}{}

\graphicspath{{images/}}
\newcommand{\pb}[1]{\vspace{0.75ex}\noindent{\bf \em #1}\hspace*{.3em}}

\usepackage{fancyhdr}
\usepackage{kantlipsum}
\fancypagestyle{plain}{
\fancyhf{}
\fancyhead[C]{Conference on \LaTeX} 

}
\usepackage{eso-pic}

\begin{document}
\AddToShipoutPictureBG*{
\AtPageUpperLeft{
\setlength\unitlength{1in}
\hspace*{\dimexpr0.5\paperwidth\relax}
\makebox(0,-0.75)[c]{\textbf{2020 IEEE/ACM International Conference on Advances in Social Networks Analysis and Mining (ASONAM)}}}}

\title{A First Look at COVID-19 Messages on WhatsApp in Pakistan}

\author{\IEEEauthorblockN{R. Tallal Javed\IEEEauthorrefmark{1},
Mirza Elaaf Shuja\IEEEauthorrefmark{1},
Muhammad Usama\IEEEauthorrefmark{1},
Junaid Qadir\IEEEauthorrefmark{1},
Waleed Iqbal\IEEEauthorrefmark{2},
Gareth Tyson\IEEEauthorrefmark{2},
Ignacio Castro\IEEEauthorrefmark{2},and
Kiran Garimella\IEEEauthorrefmark{3}
} \\
\IEEEauthorblockA{\IEEEauthorrefmark{1}Information Technology University, Punjab, Pakistan.}\\
\IEEEauthorblockA{\IEEEauthorrefmark{2}Queen Mary University of London}\\
\IEEEauthorblockA{\IEEEauthorrefmark{3}MIT}

Email: \IEEEauthorrefmark{1}(tallal.javed, msds18051, muhammad.usama, junaid.qadir)@itu.edu.pk, \IEEEauthorrefmark{2}(w.iqbal, g.tyson, i.castro)@qmul.ac.uk, \IEEEauthorrefmark{3}garimell@mit.edu
}
\maketitle
\IEEEoverridecommandlockouts
\IEEEpubid{\parbox{\columnwidth}{\vspace{8pt}
\makebox[\columnwidth][t]{IEEE/ACM ASONAM 2020, December 7-10, 2020}
\makebox[\columnwidth][t]{978-1-7281-1056-1/20/\$31.00~\copyright\space2020 IEEE} \hfill}
\hspace{\columnsep}\makebox[\columnwidth]{}}
\IEEEpubidadjcol

\begin{abstract}
The worldwide spread of COVID-19 has prompted extensive online discussions, creating an `infodemic' on social media platforms such as WhatsApp and Twitter. However, the information shared on these platforms is prone to be unreliable and/or misleading. In this paper, we present the first analysis of COVID-19 discourse on public WhatsApp groups from Pakistan. Building on a large scale annotation of thousands of messages containing text and images, we identify the main categories of discussion. We focus on COVID-19 messages and understand the different types of images/text messages being propagated. By exploring user behavior related to COVID messages, we inspect how misinformation is spread. Finally, by quantifying the flow of information across WhatsApp and Twitter, we show how information spreads across platforms and how WhatsApp acts as a source for much of the information shared on Twitter.
\end{abstract}


\section{Introduction}
\label{section: introduction}

Social media apps like Facebook, WhatsApp, and Twitter have changed the way we communicate. The information disseminated through these apps has influenced our social and cultural norms in an unprecedented way. WhatsApp is one of the most frequently used and rapidly growing social media apps in the world with more than 1.5 billion users. WhatsApp is also ranked \#1 for the average number of active users in the world per month \cite{whatsappstats}. 

WhatsApp has, therefore, become important for understanding social behavior and opinion formation. In many cases, the scope of the information shared in WhatsApp groups, is limited to a community or a country \cite{Resende2019MisInformationBrazil}. Recent studies \cite{AnalysisOfTextualMisInfoWABrazil} have shown that WhatsApp (like other social media platforms) is also used for the dissemination of misinformation. We argue that it is important to understand how this false information, either spread knowingly ("disinformation") or naively ("misinformation") influences opinion formation in different societies. This is particularly important in the global south, where despite users having low digital literacy, WhatsApp is the de facto mode through which users obtain and share information \cite{AnalysisOfTextualMisInfoWABrazil}. 


With this in mind, we perform a comprehensive analysis of COVID messages being propagated through WhatsApp in Pakistan. Pakistan is a major developing country, with approximately 37 million active social media users\footnote{\url{https://datareportal.com/reports/digital-2020-pakistan}}.
Building on the idea of how political parties around the world~\cite{reuters2019report} are using public WhatsApp groups to reach their audience, we start by monitoring a large sample of public WhatsApp groups related to politics in Pakistan. Meanwhile, a major world event occurred: On March 11, 2020, the World Health Organization (WHO) declared COVID-19 a pandemic \cite{WHO_pandemic}. Our data also holds unique value, as it encompasses non-covid groups, and gives us insight into how COVID related content, organically gets propagated across public WhatsApp groups.
Specifically, we explore the following research questions:

\begin{itemize}
    \item[\textbf{RQ1:}] What kinds of messages, about the pandemic, are being shared, on the publicly accessible WhatsApp groups of Pakistan?

    
    \item[\textbf{RQ2:}]  Is there misinformation related to COVID-19, and if so, to what extent and of which type?
    
    \item[\textbf{RQ3:}]  What is the general user behavior, and can we detect disinformation from it?
    
    \item[\textbf{RQ4:}]  What is the interplay between misinformation related to COVID-19 shared on WhatsApp and Twitter?

\end{itemize}

To explore these questions, we have collected data from 227 public WhatsApp groups starting January 10, 2020. To the best of our knowledge, this is the first dataset and analysis of COVID-19 related conversations from a country in the global south, involving multiple modalities (text and images) and multiple platforms (WhatsApp and Twitter).
We begin our investigation by analyzing the content shared in the WhatsApp groups and filtering out the COVID-19 related content. The filtered content is then further divided into text, images, videos,\footnote{In this study, we focus on text and images, leaving video analysis for future work.} and other related categories. 
Using this data, we make the following contributions:
\begin{itemize}
    \item We offer the first WhatsApp dataset consisting of discussions related to COVID-19 from Pakistan. The dataset includes texts, images and videos originating from 227 groups. The (anonymized) dataset will be made publicly available to the community.
    
    \item We show using extensive manual annotation that around 14\% of the messages related to COVID-19 had misinformation about the pandemic.
    
    \item We perform a temporal analysis of misinformation related to COVID-19 propagation, across WhatsApp and Twitter, exploring how content is copied across. 
    
\end{itemize}

\section{Related Work}

\subsection{(Mis)Information on WhatsApp}
WhatsApp has been a source of major political misinformation and propaganda campaigns~\cite{reuters2018brazil,time2019india}. 
Political parties have invested heavily in social media strategies by creating WhatsApp groups to reach WhatsApp users~\cite{goel_2018}. Surveys done in India and Brazil show that at least one in six users are part of one such public political WhatsApp group~\cite{lokniti2018,reuters2019report}. 

Garimella et al. \cite{garimella2018whatapp} provide tools to collect and analyze public WhatsApp group data at scale.
Making use of these tools, various studies have shown the extent of misinformation and manipulation on WhatsApp~\cite{evangelista2019whatsapp,resende2019mis,yadav2020understanding,garimella2020images}.
Particularly, Resende et al. \cite{resende2019mis} analyze  doctored images to fuel smear campaigns against political rivals and the dissemination of misinformation through WhatsApp groups in Brazil. 
Garimella et al.~\cite{garimella2020images} provide an analysis of image-based misinformation spread during the 2019 Indian elections and show that over 13\% of the images contained misinformation.
Melo et al. \cite{melo2019whatsapp} provide a system for gathering analyzing, and visualizing WhatsApp public group data for identification of misinformation propagated in three countries: India, Brazil and Indonesia. 
Maros et al. \cite{marosanalyzing} analyze audio messages shared on WhatsApp and characterize their propagation dynamics. The analysis is performed on 20K audio messages from 330 WhatsApp public groups and the results suggest that the audio messages with misinformation spread further more than the benign or unchecked audio messages.


\subsection{Health (Mis)Information}

A major focus of this paper is understanding the spread of health misinformation related to COVID-19. 
WhatsApp has been a major source of  health misinformation especially during the pandemic~\cite{wsj2020whatsapp}. This misinformation ranges from highlighting wrong symptoms to ineffective treatments. 
Jin et al. \cite{jin2014misinformation} reported a massive wave of misinformation on social media, especially on Twitter during the Ebola pandemic in Africa. More comprehensive details on how fake news about Ebola on social media applications is explored in \cite{fung2016ebola}.

With the ongoing surge in the COVID-19 pandemic a wealth of misinformation has already been documented. Sharma et al.~\cite{sharma2020coronavirus} provide a dashboard for analyzing misinformation about COVID-19 on Twitter. They analyze 25 million tweets and provide a country-wise sentiment analysis of how people are reacting to COVID-19. Singh et al.~\cite{singh2020first} analyze Twitter-based misinformation about COVID-19 and provide insights on how the propagation of misinformation on social media is connected to the rise in the number of COVID-19 positive cases. 
Kouzy et al. \cite{kouzy2020coronavirus} analyze Twitter-based misinformation about COVID-19 and report that tweets having the keyword ``COVID-19" contains less misinformation and tweets with keywords ``2019-ncov" and ``Corona". Cinelli et al. \cite{cinelli2020covid} provide a comprehensive analysis of the use of different social media platforms in the COVID-19 pandemic. They analyze Twitter, Instagram, YouTube, Reddit and Gab, providing a review of how the discourse on these applications is evolving. They also explore the propagation of misinformation from different questionable sources in social media.

\subsection{Our Work's Novelty}

To the best of our knowledge, there does not exist any work analyzing COVID\footnote{For brevity, we refer to COVID-19 simply as COVID and use these terms interchangably} related discussions on WhatsApp.
Since WhatsApp is arguably the most frequently used application in the world, it is important to study it to see how people are using the platform during the pandemic and how the platform facilitates the spread of COVID-19 misinformation. 
Although prior work has focused on misinformation spread via WhatsApp in Brazil and India, we are the first one to study misinformation on WhatsApp during a major pandemic. Furthermore, our analysis is focused on Pakistan, which has a thriving Muslim religious identity, which allows us to see how religion plays a role in the context of public health. 
In contrast to the majority of prior work on misinformation, which focuses on textual analysis, we also provide a detailed analysis of images related to COVID-19 and study the information spread across WhatsApp and Twitter both for text messages as well as images.





\section{Methodology}

In this section, we delineate our data collection \& annotation methodology, and discuss the related ethical issues. 

\subsection{Data Collection}
WhatsApp allows its users to create public and private groups. The public groups can be joined by any user of the platform, typically through an invite URL  of the form \url{chat.whatsapp.com/*}. These URLs  are frequently shared via other social web platforms (e.g., Facebook, Twitter) to invite third parties to join. 

\noindent\textbf{Selection of groups}.
To compile a list of relevant public groups, we looked for \url{chat.whatsapp.com} links on Facebook and Google to find group invite URLs.
We specifically targeted the popular political parties of Pakistan as these groups tend to be more active and give an idea of the political sphere. Hence, ``WhatsApp" along with political parties' names and slogans were used to search for public groups.




Based on the above parameters, we compiled a list of 282 public WhatsApp groups.
%
In order to ensure the quality of groups, we manually discarded groups that were unrelated. For instance, if a group's profile picture, group name or bio did not contain any relevant information (political aims/motivations) then it was removed. 
We further removed groups which were buying and selling things, and did not have any organic interactions/messages. This left us with 227 public groups, on which the analysis was done.

In order to find these groups, a set of queries, search engines and filters were used. These queries can be found at \url{https://cutt.ly/8yXhxBd}.
We also plan to release our anonymised dataset once the paper is accepted to encourage further research on WhatsApp data from Pakistan.

\noindent\textbf{WhatsApp data collection}.
To join and get data from the groups, we used tools provided by Garimella et al.~\cite{garimella2018whatapp}, which uses the Selenium Web Driver to automate the joining of the groups.
WhatsApp stores all message data on the user's device in an encrypted SQLite database. We used a rooted Android device to obtain the decryption key and and extracted the decrypted database every week. 
The media content, which is stored as encrypted URLs was downloaded locally and decrypted using a public tool\footnote{\url{https://github.com/ddz/whatsapp-media-decrypt}} (slightly modified for our convenience). 
WhatsApp deletes media content from their servers after a certain amount of time. As a result, when decrypting media files, we missed a small subset of the content shared (14\%). 
The joining of the groups took place over a 1 month period, as new groups were being identified.
The data collection started from 10 January 2020 onward and continued until 23 February 2020. We have complete data from all groups from the end of February until the second week of April.
The details of the dataset are summarized in Table \ref{table:dataset_overview}.

\begin{table}[hbt]
\begin{center}
\caption{Overview of our WhatsApp dataset.}
\label{table:dataset_overview}
\begin{tabular}{| r | c | c | }
\hline
\#Groups    &    227 & \\
\hline
\#Admins	& 521  & \\
\hline
\#Users    & 18,475  & \\
\hline
\#Unique users &	 16,493 & \\
\hline
Total \#messages  &    60,202 & \\
\hline

\hline
\#Text messages     &    28,497 & 47\% \\
\hline
\#Images    &    14,633 & 24.5\% \\
\hline
\#Video     &    11,196 & 18.6\% \\
\hline
\#Audio     &    2,688 & 4.5\% \\
\hline
Others      &    6,740 &   5.4\% \\
\hline
\#URLs  &    3,188 & 2.5\% \\
\hline
\end{tabular}
\end{center}
\end{table}

\pb{Twitter data.}
To compare the data we obtained from the WhatsApp groups to other open, well studied social media platforms, we also gathered data from Twitter. Specifically, we obtained historical Twitter data on an extensive list of hashtags specific to COVID-19 in Pakistan such as \#CovidPakistan, \#CoronaFreePakistan\footnote{\url{https://cutt.ly/nyXrVyp}} and other local Twitter trend variations. This gave us 800,000 tweets.

\pb{Ethics note.}
The groups joined, had been openly propagated on Facebook, Twitter, and other mediums and can be joined by anyone.
The profile bio of our WhatsApp account declares that we are collecting information for research purposes. We also anonymized the user data, before analyzing it.

\subsection{Identifying COVID-19 Text Messages}
\label{sec:text_annotation_methodology}
\begin{figure}[!hbt]
    \centering
    \includegraphics[width=0.35\textwidth]{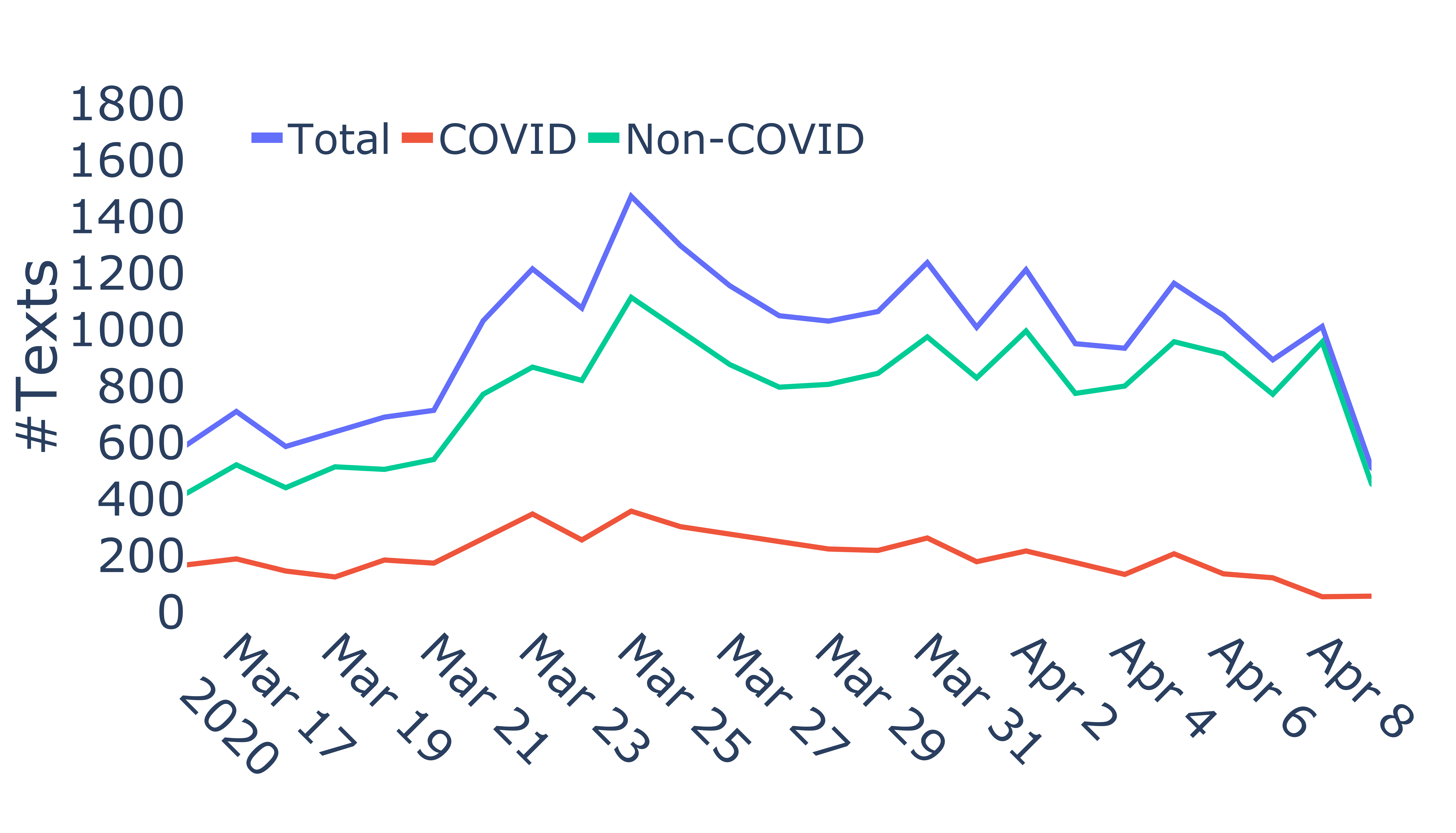}
    \caption{COVID vs. Non-COVID texts. Timeline of number of messages and messages containing COVID related keywords in our WhatsApp dataset}
    \label{fig:covid_texts_trend}
\end{figure}

We extract COVID-19 related text messages using a keyword-filtering approach. We utilize \cite{rashed2020english}, which offers a dictionary of COVID-19 \textit{English} keywords. We added small variations and multiple spellings to the dictionary to capture a wide variety of content related to the pandemic. We translated these keywords into \textit{Urdu} and used both the English and Urdu keywords to search our dataset. The final list includes keywords such as ``corona'', ``coronavirus'', ``covid-19'', ``covid'', ``covid19'', and ``corona virus'', among others.
This keyword based approach results in a high precision yet low recall method to identify COVID related messages.
Using this approach, we obtained 5,039 COVID related text messages between March 16, 2020 and April 09, 2020. 
Figure \ref{fig:covid_texts_trend} compares the number of daily COVID-19 related and Non-COVID-19 related text messages in our dataset.

\subsection{Identifying COVID-19 Images}
As we see in Table~\ref{table:dataset_overview}, around 25\% of the content is images. Hence, solely evaluating text would give a distorted view of the overall information landscape.  Naturally, image content is far harder to automatically categorize. Therefore, to extract images discussing COVID-19, manual tagging was performed

Two annotators tagged a total of 6,699 images, ranging from 16 March to 9 April, 2020. An image was declared as COVID related if it had any of the following attributes:
\begin{enumerate}
    \item Contained Coronavirus,  COVID-19,  or  any  other  related terminology in Urdu or English.
    \item Information relating to a lockdown or any restrictions being imposed/relaxed by the government on business or public/private institutions.
    \item Sharing of any precautionary measures like prayers for protection from disease, herbal medications, etc.
    \item Contained any references to the environmental or economic impact of COVID-19.
    \item Contained people with personal protective equipment, possible quarantine centers, and  people practicing or encouraging social distancing.
\end{enumerate}

\begin{figure}[!hbt]
    \centering
    \includegraphics[width=0.35\textwidth]{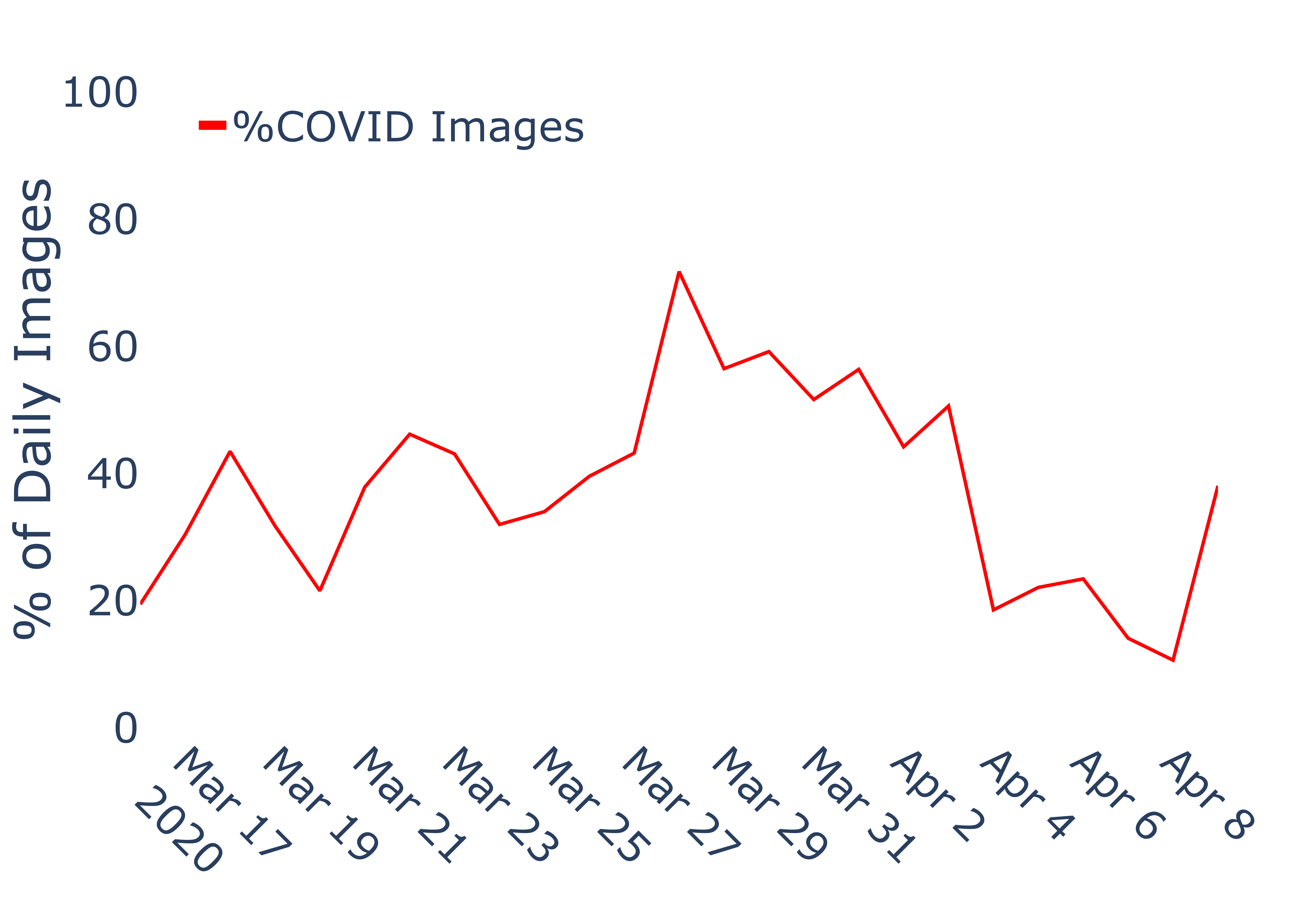}
    \caption{Timeline of the percentage of images and images containing COVID related content in our WhatsApp dataset}
    \label{fig:covid_images_trend}
\end{figure}

An inter annotator agreement score of 98\% was observed. In cases of a conflict, the annotators were allowed to mutually discuss and agree upon a label. A total of 2,309 (34.5\%) images were identified as COVID related, while 4,390 (65\%) were identified as non-COVID images. 
For context, Figure \ref{fig:covid_images_trend} shows the percentage of COVID related images over time. We see that, as the pandemic intensifies, so does the fraction of related images.


\section{RQ1: Information Sharing on COVID-19}
\label{subsec:covid_content_classification}
In this section, we answer our first research question, to understand what types of COVID related information is being shared on WhatsApp.
We first annotate the data into 5 overalapping categories and then use these categories to understand the types of information.

\subsection{Message Type Categorization}
In order to further characterize the types of COVID-19 related content being shared within our WhatsApp groups, we categorized the COVID related text and images manually into the five categories: \textit{Information}, \textit{Misinformation}, \textit{Jokes/Satire}, \textit{Religious}, and \textit{Ambiguous}. The categories were chosen based on a preliminary manual exploration of the COVID-19 content. The categories are not mutually exclusive and as a result, a single message (text/image) can belong to multiple categories. We describe each category below:

\begin{enumerate}
\item \textbf{Information:}
    This category consists of WhatsApp content that contains either factual News or COVID related facts. News reports are fact checked using Poynter's COVID Facts database\footnote{\url{https://www.poynter.org/ifcn-covid-19-misinformation/}} which contains all of the falsehoods detected by a large number of fact checking organizations. In addition, AFP Pakistan Fact Check\footnote{\url{https://factcheck.afp.com/afp-pakistan}} is used to verify news articles. The contents of the text or image are evaluated against the falsehoods in the database to verify their validity.
    Google search was also used to verify certain claims not present in the Poynter dataset. If the news is reported by a reputed news source, then it is labelled as ``Information''. A news source is considered reputed if it has a satellite news channel or newspaper at a national level. COVID related facts are verified using WHO's COVID Information and pervalent myths.\footnote{\url{https://www.who.int/emergencies/diseases/novel-coronavirus-2019/advice-for-public/myth-busters}}
    
    \item \textbf{Misinformation:}
    This category is the inverse of the above `Information' category. Any content which is either verified to be misinformation or could not be verified as credible information is placed in this category. Content was checked using Poynter COVID-Facts and Falsehoods database, AFP Pakistan and WHO's COVID Informations and COVID Myths.
    
    \item \textbf{Jokes/Satire:}
    This class consists of content that intends to poke fun at the COVID-19 pandemic itself or any COVID related government/political actions using sarcasm, satire or memes. It also contains content that consists of non-factual opinions/analysis regarding current COVID related events or government actions.
    
    \item \textbf{Religious:}
    Since Pakistan has a 98\% Muslim population, religion plays an important role in information dissemination. A religious theme in content is identified by looking for (i) references to spiritual texts, (ii) quotes of religious scholars (called \textit{Maulana}, \textit{Mufti}, or \textit{Sheikh}), and (iii) mentions of religious acts such as prayer, fasting etc. 
    
    \item \textbf{Ambiguous:}
    If the content does not have enough information to be classified into one or more of the above categories, it is then assigned to the `Ambiguous' category. This category mainly consists of content where people are distributing Personnel Protective Equipment (PPE), social media requests to follow/subscribe, contact information of NGO's, donation requests, etc.
\end{enumerate}


\begin{figure}[!hbt]
    \centering
    \includegraphics[width=8cm]{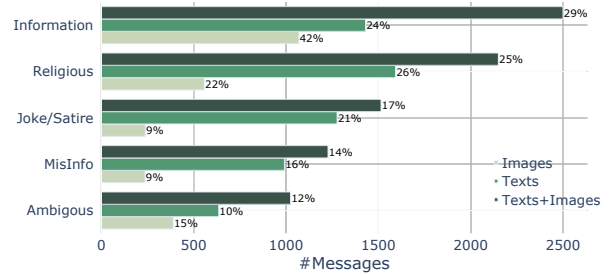}
    \caption{Percentage of COVID-19 images (Light Green), texts (Medium  Green) and texts + images (Dark  Green) for each category. Notably, 14\% of the total messages were labeled to be misinformation.}
    \label{fig:wa_labelled_texts_images}
\end{figure}

Note that to maintain consistency in annotation quality between images and text (images were annotated by two annotators whereas text was done by only one annotator), we randomly sampled 25\% of the 5,039 COVID-19 related texts, and validated the annotation. We showed the 25\% random sample of messages to one additional annotator and measured the agreement with the original annotator. We find an 82\% agreement between the two annotators, with one or more common labels counted as an agreement for our non-mutually exclusive classes. The majority of disagreements were between Information, Jokes/Satire and Religious classes. This is because a lot of texts contain different proportions of the three. Very few disagreements were observed when one of the annotators tagged a text as Misinformation, which were resolved after a discussion between the two annotators.

\subsection{Message Type Analysis}


We now analyze the different types of COVID-19 related content, in both texts and images on WhatsApp, based on the above annotation. 
We have a total of 5,039 texts and 2,309 images which discuss COVID related information between March 16 and April 9, 2020. The overall distribution of texts and images into the COVID-19 related content categories are shown in Figure \ref{fig:wa_labelled_texts_images}. A majority of the content is simple information (29\%), containing news articles, latest government actions and health information related to COVID-19. This is followed by religious content (25\%). The large amount of religious content emphasizes the importance of religious sentiment within the society, especially during times of uncertainty created by the pandemic. Religious scholars and Holy Verses from religious books were cited in these messages. Religious content was also event focused. For instance, a ban on congregational prayers and the rigorous COVID testing of a group of religious people on a proselytising trip resulted in messages criticising these government actions. 

This was followed by Jokes/Satire representing 17\% of the messages. Political actions by rival parties, and government officials were frequently ridiculed and mocked, including personally targeted attacks. For instance, many of such texts were against government action of opening the border with Iran, blaming the officials for bringing Corona to Pakistan. Interestingly, a non trivial amount (14\%) was Misinformation, which shows that one in seven messages shared contained some misleading information. This included fake news reporting deaths of politicians, fake quotes from famous personalities and international figures, or fake COVID origin stories. We show detailed analysis of misinformation in Section~\ref{sec:misinfo_analysis}. 

Finally, a small fraction of messages (12\%) were labelled ``Ambigious''. Among these, a significant portion of texts contained Facebook and YouTube follow/subscribe requests to COVID related pages and channels, donation requests, or shared contact information for COVID-affected and poverty-stricken people. We also found some images depicting quarantine centers, hospitals, doctors, and patients, which did not fit into the above categories due to the lack of context.




\subsection{Lifetime of Messages}
In this section, we try to understand the \textit{impact} of the various types of COVID messages on WhatsApp. 
To do so, we analyzed the lifetime of various types of COVID related messages. The lifetime of a message is the difference between the last and first appearance of a message (in hours) in our dataset. First, we grouped together perceptually similar images using a popular, state of the art image hashing tool from Facebook known as PDQ hashing.\footnote{\url{https://github.com/facebook/ThreatExchange}}
The hashes were generated for all the COVID related images and then instances of similar images were clustered together by using Hamming distance, with a threshold of 70\%, between the hashes.
The difference between first and last appearance of a representative image in each cluster was considered as the lifetime of that image. For texts, exact string matching was used to find the first and last appearance of a text. Table \ref{table:wa_covid_temporal} shows the mean lifetime of messages belonging to the various COVID-19 content categories.

Each category exhibits distinct mean and variance measure for lifetime. The most short lived messages belong to the ``Jokes/Satire'' category. This appears logically coherent since jokes, opinions and satirical texts are generally dictated by events and die out quickly as the public focus shifts from one event to another. 
Interestingly, the lifetime of a message containing misinformation is quite high, for both text (7 hrs) and images (5.5 hrs), especially compared to the Information category. This means that misinformation tends to persist longer compared to information, which supports existing studies showing similar results~\cite{vosoughi2018spread,garimella2020images}. Given that WhatsApp is a closed platform with no content moderation or third party fact checking, the fact that misinformation tends to stick around longer might be expected, but might also  be problematic when compared to other social networks, where eventually corrections can be issued.


\begin{table}[!htb]
\begin{center}
\caption{Lifetimes of COVID-19 related texts and images shared on WhatsApp. \emph{Misinformation tends to have the highest mean lifetime.}}
\label{table:wa_covid_temporal}
\begin{tabular}{| l | c | c | c | c | c | c | }
\hline
\multicolumn{4}{| c |}{\textbf{Text Messages}}\\
\hline
\textbf{Label}  & \textbf{Num.}  &   \textbf{Mean}    &   \textbf{Std Dev}    \\
                &  \textbf{texts}  &   \textbf{(hrs)}    &   \textbf{(hrs)}  \\
\hline
\textit{Information}   &   1108  & 2.75     &   21.98  \\
\hline
\textit{Religious}  & 829 &   6.98       &   29.15  \\
\hline
\textit{Jokes/Satire}  & 919 &   1.92     &   9.15  \\
\hline
\textit{Misinformation}  & 596 &   7.0    &   28.03  \\
\hline
\textit{Ambiguous}  & 313 &   10.05    &   39.2  \\
\hline
\multicolumn{4}{| c |}{\textbf{ Images}}\\
\hline
\textbf{Label}  & \textbf{Num.}  &   \textbf{Mean}     &   \textbf{Std Dev}    \\
                &  \textbf{images}  &   \textbf{(hrs)}    &   \textbf{(hrs)}  \\
\hline
\textit{Information}   &   1069  & 0.55     &   2.87  \\
\hline
\textit{Religious}  & 557 &   2.70       &   6.14  \\
\hline
\textit{Jokes/Satire}  & 238 &  1.21      &   4.07  \\
\hline
\textit{Misinformation}  & 236 &   5.57    &   9.17  \\
\hline
\textit{Ambiguous}   &  389   & 1.35     &   4.31  \\
\hline
\end{tabular}

\end{center}
\end{table}
\begin{figure}[!hbt]
    \centering
    \includegraphics[width=8cm]{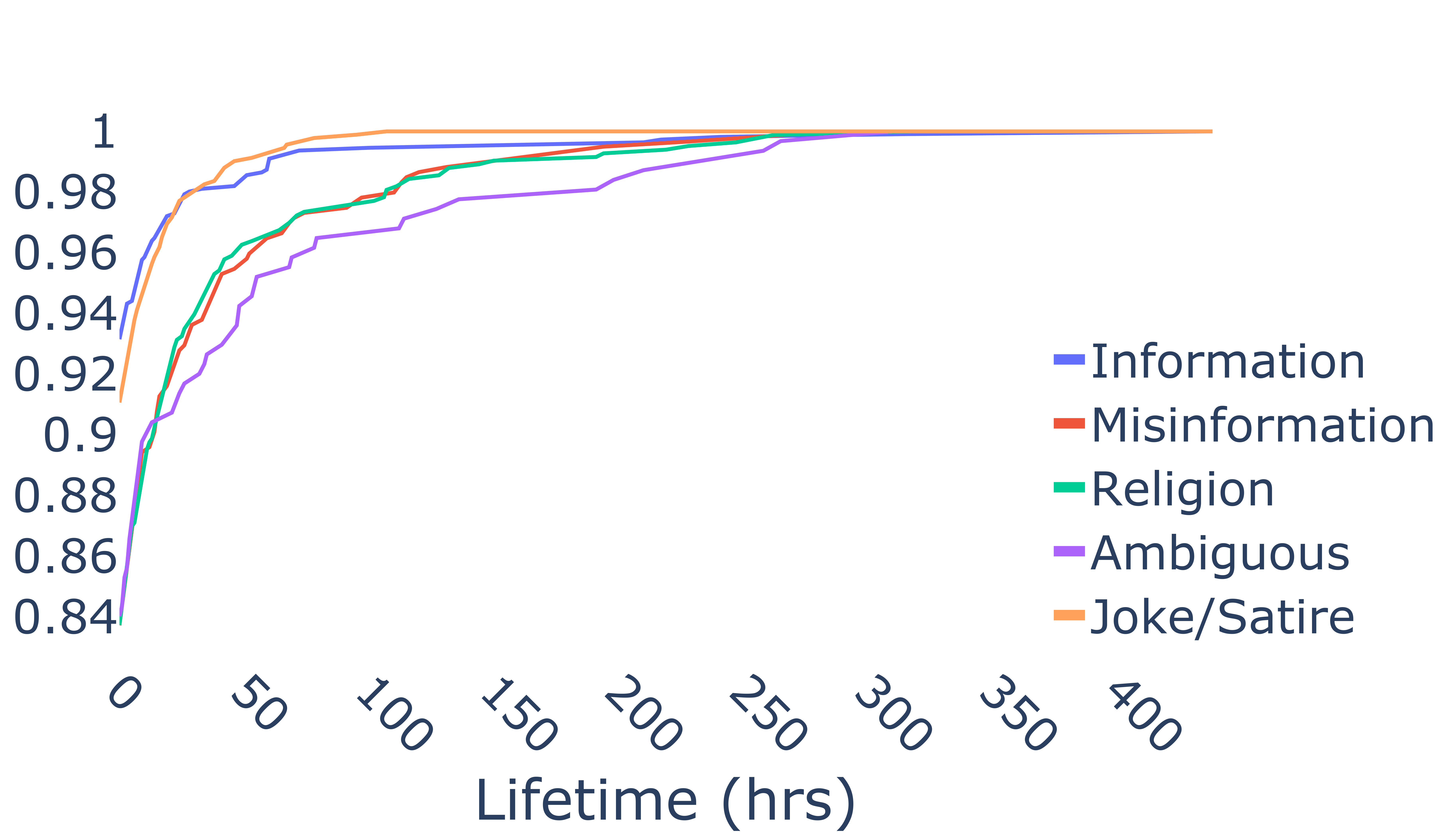}
    \caption{CDFs of Text of COVID-19 categories (Note the broken y-axis: For better Interpretability).}
    \label{fig:cdf_texts}
\end{figure}

\section{RQ2: COVID-19 Misinformation}
\label{sec:misinfo_analysis}
In this section, we specifically look at misinformation posts, and characterize the types of misinformation shared on WhatsApp.
We first categorize the types of misinformation based on reports from popular fact checking organizations and then base our characterization on this categorization.

\subsection{Misinformation Message Analysis}

\begin{figure}[!hbt]
    \centering
    \includegraphics[width=9cm]{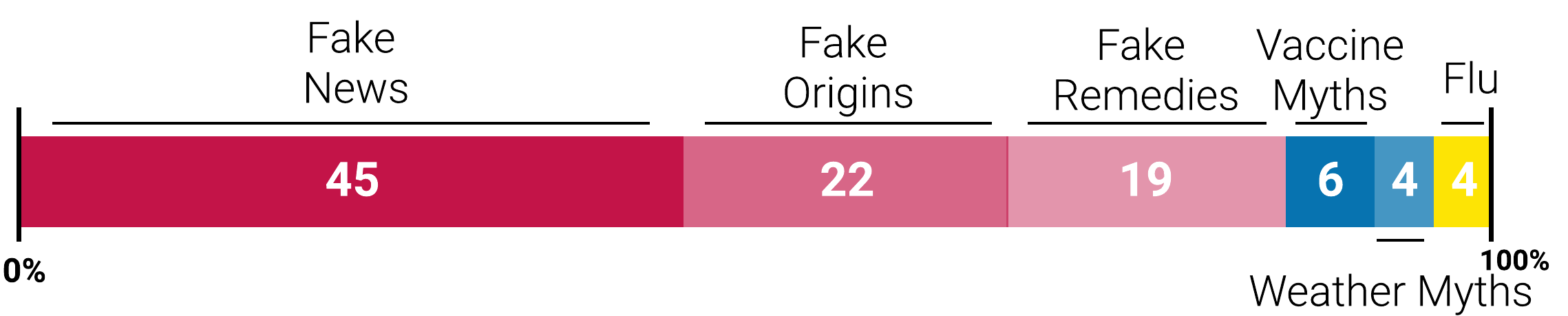}
    \caption{Percentage of texts on WhatsApp for each type of COVID-19 related misinformation.}
    \label{fig:wa_text_misinfo_bar}
\end{figure}

The distribution of the various categories is shown in Figure~\ref{fig:wa_text_misinfo_bar}.

\pb{Fake News.} The most frequent form of COVID related misinformation is in the form of fake news with 45\% of misinformation texts. This includes fake news pertaining to COVID positive tests and COVID related deaths of world figures such as Ivanka Trump, Prince Williams and even the current Prime Minister of Pakistan, Imran Khan. 
Conspiracy theories about Bill Gates intending to place RFID chips in people to track COVID-19 were also seen. Ironically, fake news were also observed regarding a doctored government action announcing `Punishment for Spreading Fake News on social media'.

\pb{Fake Origins.} The second most prevalent form of COVID-19 related misinformation is claiming fake origin stories for the virus with 22\% of the misinformation texts. Fake origin stores include a Corona named lake in Kazakhstan from which the virus came to being grown in a lab in China or United States. A few Hollywood movies, namely `Contagion', `Resident Evil' and `I am Legend' along with the book `The Eye of Darkness' were frequently mentioned stating that COVID-19 had been predicted by them.

\pb{Fake Remedies.} Making up roughly 20\% of the misinformation, this type contains bogus remedies and treatments such as the 1-minute breath hold test to detect COVID, and various items like basil seeds, gargling with salt or garlic water, honey lemon tea and even Hepatitis-C medicine as cures to COVID-19.

\pb{Vaccine Myths.} The fake origin stories were sometimes accompanied with claims of the vaccine already being developed and being used as an economic leverage. Countries such as Israel, China and United States were mentioned with claims of the vaccine already developed. This category makes up around 6\% of the misinformation texts.

\pb{Weather Myths.} Four percent of the misinformation claims that the virus can not survive in winter, summer or rainy seasons and the outbreak would die down on its own.

\pb{Flu Comparison.} Only 2\% of the misinformation attempted to downplay the symptoms and severity of the disease by comparison to the common seasonal flu. 
Even though this narrative was popular elsewhere (e.g. US), it did not have much salience in Pakistan, with the general public acknowledging COVID-19 as a distinct and more dangerous disease as compared to the common flu.

\subsection{Lifetime of Misinformation}

The temporal properties of the various categories of misinformation are analyzed in table \ref{table:wa_covid_misinfo_temporal}.
For each message, we compute the lifetime as the difference between its last and first occurance. 
The `Fake News' category has the shortest lifespan as evidenced by the lowest mean of 4 hrs. This seems to be consistent with the hypothesis that event triggered content is short-lived, with similar  properties to the `Jokes/Satire' category of COVID-19 related textual content. 
The highest lifespan is for the `Fake Remedies' category with a mean life of 10 hrs, which is significantly larger than other major categories. This indicates that content that is not tethered to a social event is more likely to being in circulation on a social media platform like WhatsApp.

\begin{table}[!htb]
\begin{center}
\caption{Lifetime of misinformation texts shared on WhatsApp. } \label{table:wa_covid_misinfo_temporal}
\begin{tabular}{| l | c | c | c | }
\hline
\textbf{Label}  & \textbf{Num.}  &   \textbf{Mean}   &   \textbf{Std Dev}     \\
                & \textbf{texts}   &   \textbf{(hrs)}  &   \textbf{(hrs)}      \\
\hline
\textit{Fake News}  & 307 &   4.06    &   18.7  \\
\hline
\textit{Fake Origins}  & 171 &   9.4    &   35.3  \\
\hline
\textit{Fake Remedies}  & 125 &   10.6    &   33.8   \\
\hline
\textit{Weather Myths}  & 26 &   27.57    &   67.6   \\
\hline
\textit{Flu Comparison}  & 16 &   6.68    &   16.66  \\
\hline
\end{tabular}
\end{center}
\end{table}
\vspace{-0.2in}






\section{RQ3:User Behaviour Analysis}
Every WhatsApp group has two types of users: (1) \textit{producers} and (2) \textit{consumers}. Some users share and post messages whereas others silently observe. In general, producers are few and consumers are many (Table \ref{table:dataset_overview}). In this section, by examining the user behavior, we hope to understand if there is any deliberate spread of disinformation. 

\begin{figure}[!hbt]
    \centering
    \includegraphics[width=8cm]{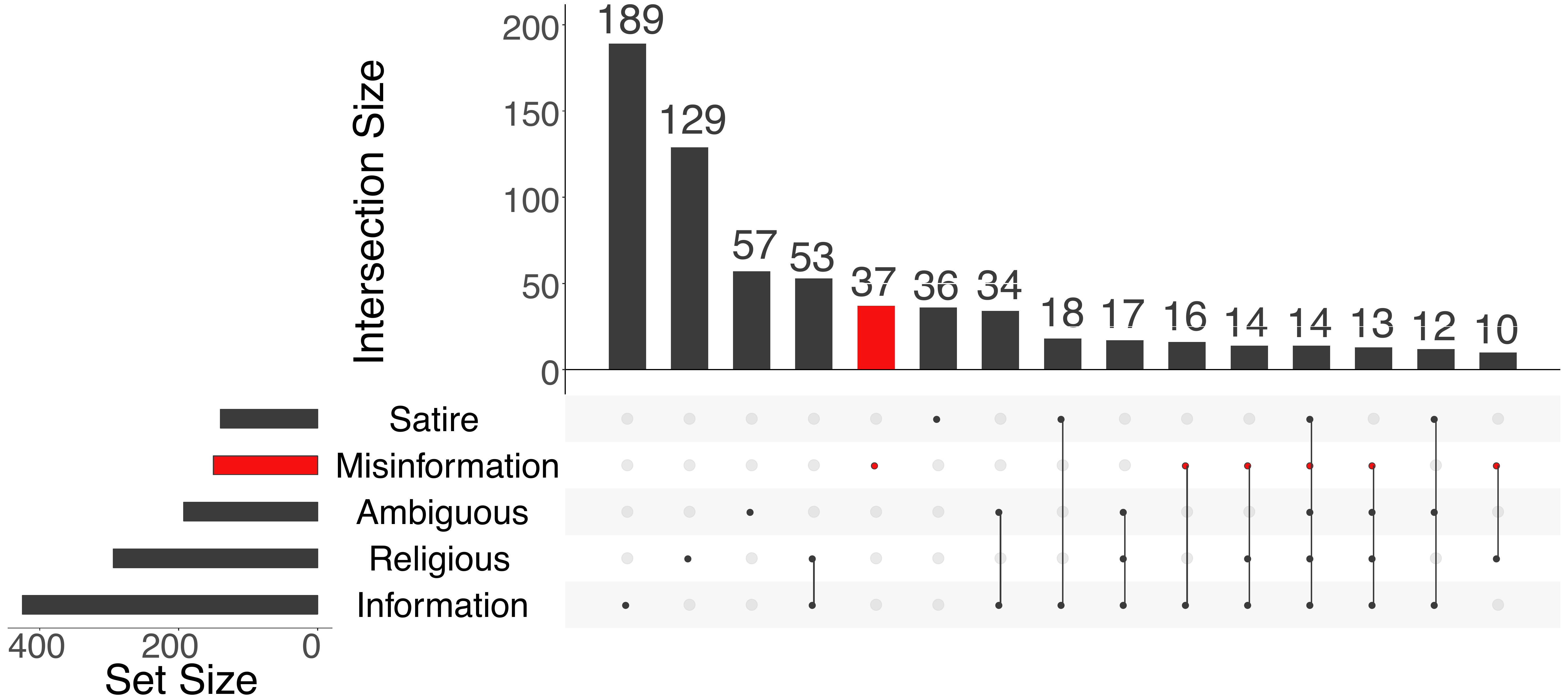}
    \caption{UpSet plot for users posting COVID related images. Only the top 15 intersection sets are visualized. \emph{A lot of users are sharing information and religious content, whereas some share misinformation.}}
    \label{fig:upset_user_images_behavior}
\end{figure}

We use ``UpSet''\footnote{For an introduction, see \url{https://www.ncbi.nlm.nih.gov/pmc/articles/PMC4720993/}.} plots, to visualize user behavior, where every set is a unique user. The bottom matrix (combination matrix) of Figure~\ref{fig:text_upset},\ref{fig:upset_user_images_behavior} shows the intersections of the sets across COVID categories, while the bars on the top indicate the number of users (sets) within that intersection. The bars on the left indicate total users (sets) within a given category.

\subsection{Text Sharing Trends}
\begin{figure}[!hbt]
    \centering
    \includegraphics[width=8cm]{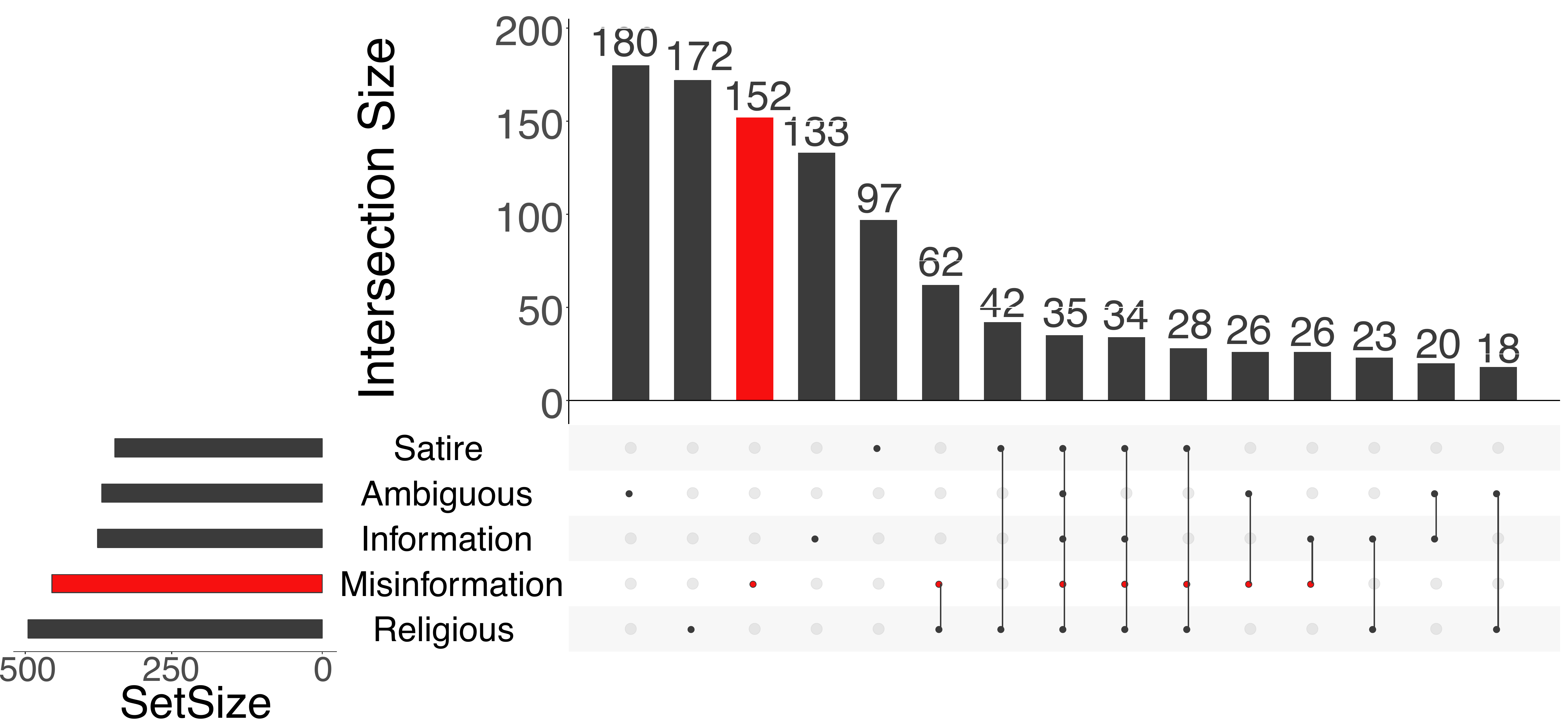}
    \caption{UpSet plot for users posting COVID related texts. Top 15 intersection sets are visualized. \emph{More users appear to share texts that belong to a single category.}
    \label{fig:text_upset}}
\end{figure}

The UpSet plot in Figure~\ref{fig:text_upset} is plotted against the text messages shared by individual users.
If we observe the combination matrix, we observe that users are sharing textual content belonging to single category. The most exclusively shared category is ``Ambigous''. It can be attributed to the users that join WhatsApp groups intending to share advertisements and call-for-donations only. The second and third highest intersection sets are for ``Religious'' and ``Misinformation'' being shared exclusively. This deviates from the trend observed for Images. People are more likely to \textbf{exclusively} share ``Texts'' containing ``Misinformation'' as compared to ``Images'' containing ``Misinformation''. This prompts for need for more research in finding traces of disinformation within text messages.

\subsection{Image Sharing Trends}
The UpSet plot shown in Figure~\ref{fig:upset_user_images_behavior} is made against the type of Images individual users are sharing. It is good to see that the majority of users are sharing correct information about the pandemic. An encouraging trend is that users are not exclusively sharing misinformation, rather a mix of content is being shared. Only 37 users exclusively shared misinformation, whereas 67 users shared a mix of content, along with misinformation. 

To further understand if there is disinformation, we tried to see if a specific type of image is being spread more than others. Using the clusters we already had, made via PDQ hashing and Hamming distance, clusters having misinformation were identified. The number of images within these clusters are a good indicator of the impact of given misinformation on the network. 23 unique images were shared more than 1 times and only 8 were shared more than 5 times.

This implies that even if we consider the images shared multiple times to be disinformation, the quantity of disinformation is very low. Hence we believe, the misinformation is, rather than being an organized effort, is mainly being spread due to lack of awareness.

\section{RQ4: Cross Network Information Spread}

Finally, we answer how information flows between WhatsApp and Twitter. Given that each social network has different properties (closed vs. open), affordances (e.g.,the ability to see how popular a content is with retweet/like count) and user bases, such a comparison is interesting.

\subsection{Methodology}

Twitter was chosen as most of Twitter data is public, and it serves as one of the major information conduits. We obtained more than 0.8 million unique tweets starting from January 10 to April 9 using an exhaustive list of hashtags related to COVID-19 in Pakistan. It must be noted, that both Twitter and WhatsApp datasets only represent a subset of the actual activity and in no way can be thought to represent the full networks' behaviour. In order to understand the flow of information, images and text present within WhatsApp, from 16 March to 9 April 2020, were compared with tweets in the same time range. To find similar content across Twitter and WhatsApp, PDQ hashing was used for images and fuzzy string matching was used for text messages.


\subsection{Cross Platform Image Spread}
\begin{figure}[t]
        \includegraphics[width=0.9\linewidth, height=5cm]{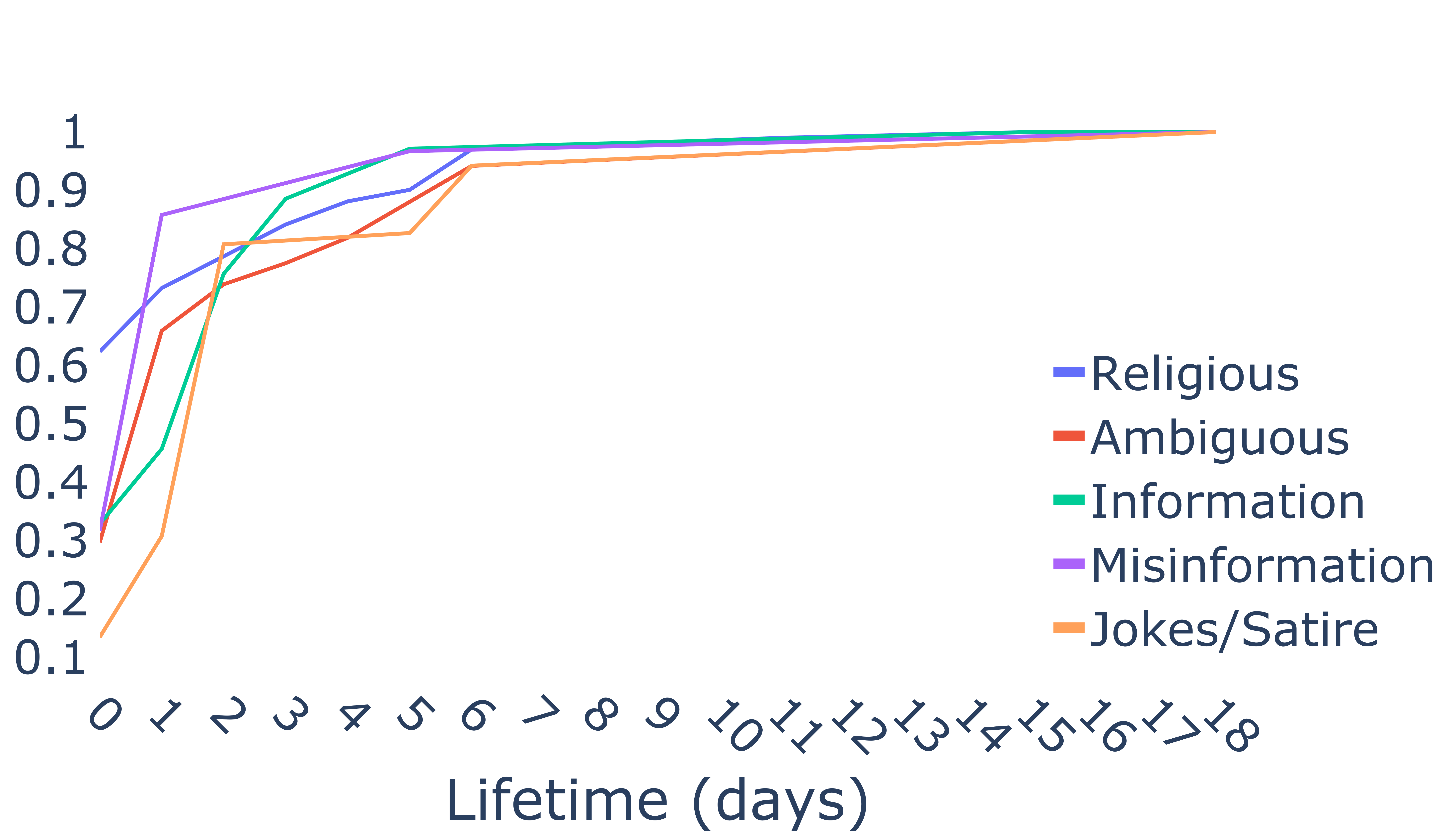} 
        \caption{CDFs of life of an image, along with content type, as seen on Twitter. Twitter tends to hold a message alive for a coupe of days. \emph{A healthy trend is that Information tends to live the longest on Twitter}.}
        \label{fig:cdf_all_twitter}
    
    
\end{figure}

To understand how images were propagated across networks, we isolate the tweets containing image content from the Twitter dataset. This covers a total of 67,119 images. We then generate PDQ hashes for both WhatsApp and Twitter images, and matching two images if their hashes have a Hamming distance of 40 (default value suggested by PDQ). Around 1,500 similar images were found common to both WhatsApp and Twitter, within the date range of interest (16 March and 9 April).
Out of these 1,500 images, 541 were COVID-19 related.

Table~\ref{table:twitter_image_clsasification} breaks down the images into the categories previously described, alongside the average number of retweets, replies and the lifetime of the image (difference between first and last appearance of an image on Twitter). Firstly, we observe that the largest category of images shared across both Twitter and WhatsApp is that of misinformation (29\%). We can also see that misinformation tweets have a high average number of retweets, potentially reaching tens of thousands of users. This is signal in Twitter (or other social networks) which does not exist on WhatsApp, where the social popularity signals like retweet or like counts are shown.

Compared to the other categories, misinformation on Twitter tends to die quicker. Figure~\ref{fig:cdf_all_twitter} shows a CDF of the lifespan of the various categories of images on Twitter. We see that most types last at most a day, with most having a long tail lasting weeks. This is very different from the pattern we observed on WhatsApp, where on average, most images lasted only a few hours. Images containing information have a much higher lifetime, which is also in stark contrast to what we observed on WhatsApp. 

The difference in longevity of a message, as seen in Table \ref{table:wa_covid_temporal} and \ref{table:twitter_image_clsasification}, points towards the nature of interactions. Interactions on WhatsApp are immediate, as new messages constantly replace old ones. Whereas on Twitter old tweets can be easily brought back into limelight, using retweets, comments and likes by influential people. Hence the ability of Twitter to keep conversation, around a tweet, alive for a long time could be the reason for the overall life of COVID tweets (in days) compared to WhatsApp (in hours).

\begin{table}[ht]
\begin{center}
\caption{Characteristics of images mapped between Twitter and WhatsApp.} 
\label{table:twitter_image_clsasification}
\begin{tabular}{ | c | c | c | c | c | }
\hline
\textbf{Label}	&	\textbf{Num.}	&	\textbf{Retweets}	&	\textbf{Replies} 	&	\textbf{Life}	\\
& \textbf{images} & (Mean) & (Mean) & (days) \\
\hline
\textit{Information}    & 79    &  64.18   &   75.27	&   5.05    \\
\hline
\textit{Religious}   &   108     &   42.66    &   45.82    &   3.25    \\
\hline
\textit{Jokes/Sarcasm}    & 104     &   26.96   &   89.71  &   3.0 \\
\hline
\textit{Misinformation}  &   183 &   67.72   &   444.28   &   1.6 \\
\hline
\textit{Ambiguous}    & 146     &   156.82  &   309.64	    &	4.2 \\
\hline
\end{tabular}
\end{center}
\end{table}

\begin{figure}[!ht]
    \centering
    \includegraphics[width=0.5\textwidth]{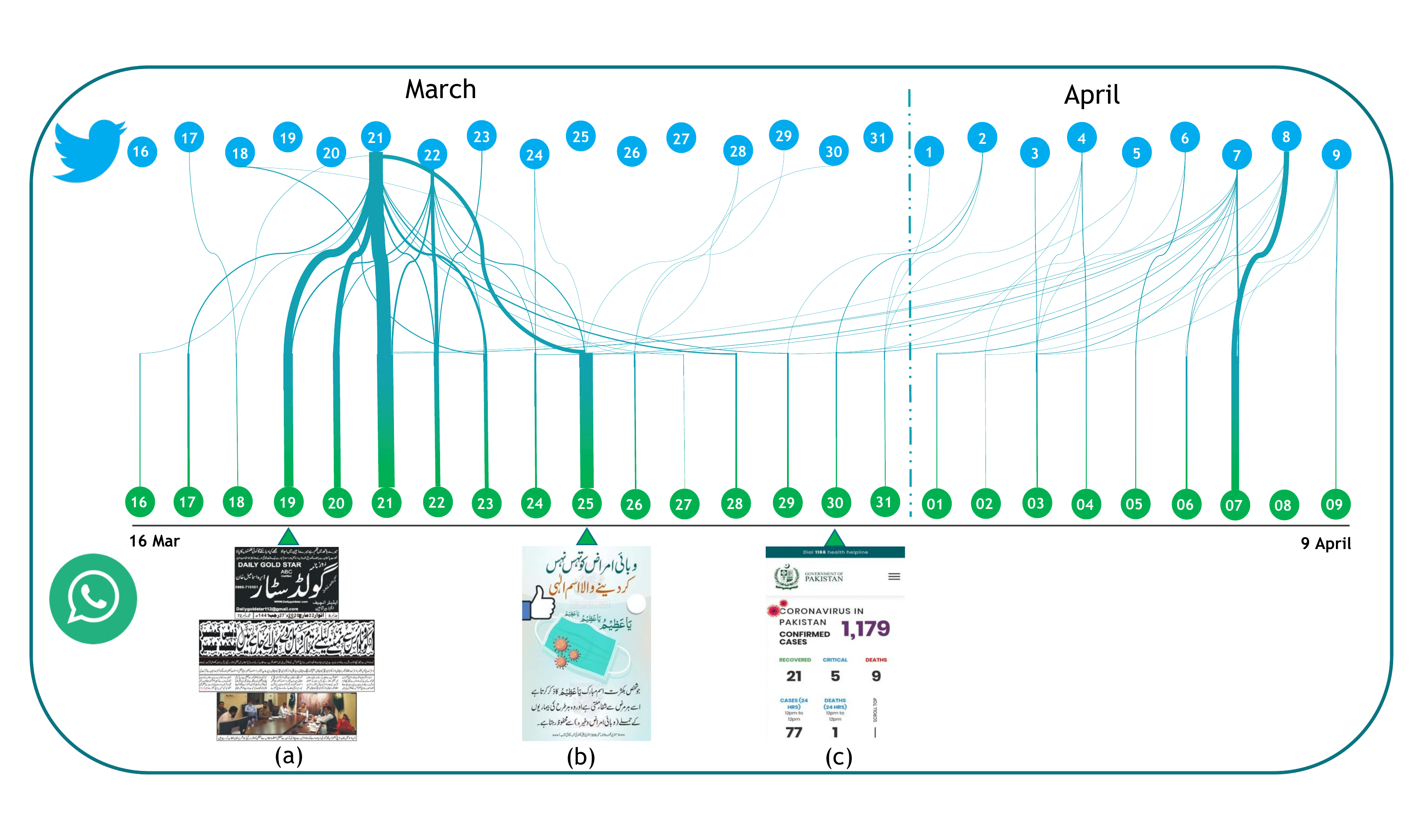}
    \caption{COVID images' temporal flow across WhatsApp and Twitter (a line's thickness depicts the number of images flowing across). \textit{Some Observations}: \textit{a)} a news snippet originates from WhatsApp on 19th March and is seen on Twitter on 21st; \textit{b)} religious supplication to fight COVID is observed on WhatsApp 2 days earlier than on Twitter; \textit{c)} official stats of COVID patients seen on 30th March on WhatsApp earlier than on Twitter.}
    \label{fig:flow-timeline}
\end{figure}

A schematic diagram of the temporal flow of images across WhatsApp and Twitter is shown in Figure~\ref{fig:flow-timeline}. The Figure also provides three example case studies of images which originated on WhatsApp and went on to become widely retweeted on Twitter. From our analysis of the timelines of images observed on both platforms, we can conclude that most of the images are seen originating from WhatsApp and then appearing on Twitter (in our dataset). On average an image appears on WhatsApp 4 days earlier as compared to Twitter. As a result in light of the data analyzed, it can be concluded that WhatsApp plays a critical role in COVID related content dissemination to other networks in Pakistan. This is especially important in the context of results from Table~\ref{table:twitter_image_clsasification} with a majority (29\%) of the content that is common between the two platforms being misinformation, compared to only 12\% being information. 

\section{Conclusions}
In this paper, we have provided the first detailed analysis of Pakistani WhatsApp public groups, focusing on the COVID-19 discourse. We have analyzed WhatsApp text and image messages collected for more than 6 weeks from 227 public WhatsApp groups to shed light on the salient misinformation dissemination trends and to share insights on how Pakistani social media users are experiencing and responding to the COVID-19 pandemic. Our work is unique as this is the first work to not only study misinformation trends on WhatsApp but also find a relation between WhatsApp and Twitter. Our analyses showed that while it is true that the majority of shared information is not misinformation, misinformation seems to have a longer lifespan on WhatsApp compared to other types of COVID messages (the lifetime of misinformation is roughly 4 times that of correct information). On Twitter the inverse was seen, as COVID misinformation tended to disappear from Twitter 3 times faster than information.  This can potentially be attributed to the open nature of Twitter, and how a vast number of users can publicly negate such tweets. While observing user behavior, we found 8 images that could be attributed to organized disinformation, other than that, we did not find any evidence of disinformation within images. Whereas more work is required in detecting disinformation via text messages. We conclude by saying that our dataset has only scratched the surface of how user interactions happen on WhatsApp. More work needs to be performed to understand user behavior, and new ways need to be proposed to detect misinformation in such closed networks.

{
\bibliographystyle{IEEEtran}
\bibliography{ref.bib}

}
\end{document}